\documentclass[pre,twocolumn,showpacs]{revtex4}
\usepackage{epsf}
\begin{document}
\title{
Exact Solution for the Influence of Spectral Diffusion
on Single-Molecule Photon-Statistics} 
\author{Y. He, E. Barkai}
\affiliation{Department of Chemistry and Biochemistry, Notre Dame University, Notre Dame, IN 46556.}

\date{\today}

\begin{abstract}

 We investigate the distribution of number of photons emitted
by a single molecule undergoing a spectral diffusion process
and interacting with a continuous wave field. 
Using a generating function formalism
an exact analytical formula for Mandel's $Q$ parameter is obtained.
The solution 
exhibits transitions between: (i)  Quantum sub-Poissonian
and Classical super-Poissonian behaviors,
and (ii) fast to slow modulation limits.
Our solution yields the conditions on the magnitude
of the spectral diffusion
time scales  on which these transitions 
are observed. We show how to choose the Rabi frequency
in such a way that the Quantum sub-Poissonian
nature of the emission process becomes strongest,
we find ${\Omega^2 \over \Gamma} = {\Gamma_{{\rm SD}} + \Gamma \over 2}$
where $\Gamma_{{\rm SD}}$ ($\Gamma$) is the spectral diffusion (radiative)
contribution to the width of the line shape. 

\end{abstract}

\pacs{82.37.-j, 05.10.Gg, 33.80.-b, 42.50.Ar}

\maketitle

Physical, Chemical, and
Biological systems are investigated in many laboratories
using single molecule
spectroscopy  \cite{MO}. 
The investigation
of the distribution of number of photons emitted
from a single molecule source is the topic of extensive theoretical research  
e.g. \cite{MM,Schenter,Berez,Barsegov2}.
An important mechanism responsible for fluctuations 
in photon statistics from single molecule
sources is spectral diffusion. In many cases
the absorption frequency
of the molecule will randomly change due to different
types of interactions between the molecule and
its environment (e.g. \cite{MO,Geva,PRL,ADV} and Ref. therein).
For example in low temperature glasses flipping two
level systems embedded in the glassy environment, 
induce spectral jumps in the absorption
frequency of the single molecule under investigation \cite{Geva}. 
In this way the molecule may come in and out of resonance with
the continuous wave laser field with which it is interacting.

 Obviously a second mechanism responsible for fluctuations
of photon counts is the quantum behavior of the 
spontaneous emission process \cite{Wolf,Plenio}. 
 In his fundamental work Mandel \cite{Mandel1} showed that a single
atom in the process of resonance fluorescence, {\em  in the absence of
spectral diffusion}, 
exhibits sub-Poissonian photon statistics \cite{Short}.
Photon statistics is characterized by
Mandel's $Q$ parameter
\begin{equation}
Q= {  \overline{N^2}  - \overline{N }^2 \over \overline{N}} -1
\end{equation}
where $N$ is the number of emitted photons within a certain time 
interval. The case $Q<0$ is called sub-Poissonian behavior,
while $Q>0$ is called super-Poissonian. Sub-Poissonian statistics
has no classical analog \cite{Mandel1}. 
Briefly, the effect is
related to anti-bunching of photons emitted from a single source
and to Rabi-oscillations of the excited state population
which favors an emission process
with some periodicity in time. 
Sub-Poissonian statistics and anti-bunching were detected
in several single molecule experiments, for example  \cite{Or,Zu1}.

 In this Letter we  
obtain an exact analytical
expression for the $Q$ parameter in the long time limit, for a single molecule
undergoing a stochastic spectral diffusion process. 
For that aim we use the powerful generating function method of Zheng and
Brown \cite{Brown}. If the spectral diffusion process is
slow enough super-Poissonian behavior of $Q$ is expected (see details
in text).
Our analytical expressions classify the
transitions between sub and super Poissonian statistics.
They  give the conditions
on the spectral diffusion time scale
and magnitude of spectral jumps for sub-Poissonian non-classical
behavior to be observed.
Our result is valid for weak
and strong excitation (i.e arbitrary Rabi frequency).
It yields the lower bound on $Q$.
The solution shows how in experiment we may choose the Rabi
frequency so that the quantum nature
of the photon emission process becomes larger,
namely how to minimize $Q$ in the Sub-Poissonian
regime. This is important for the efficient
detection of quantum effects
in single molecule spectroscopy, since choosing too small or
too large values of the Rabi frequency results in very small
and hence undetectable values of $Q$.
Finally our result is used to test the 
semi-classical linear response theory (i.e., weak Rabi frequency)
of 
Barkai, Jung, and Silbey
\cite{PRL,ADV}.  
The semi-classical theory yields $Q>0$, while the main focus of
this manuscript is on the quantum regime $Q<0$.

 Our starting point are the Zheng--Brown \cite{Brown}
generalized optical Bloch
equations
 describing a chromophore with single excited
and ground state
\begin{equation}
\begin{array}{l}
 \dot{{ \cal U}}\left(s\right) = - {\Gamma \over 2} {\cal U}\left(s\right) 
+ \delta_L(t) {\cal V}\left( s \right) \\ 
 \dot{{\cal V}}\left(s\right) = -  
\delta_L(t) {\cal U}\left( s \right)-{\Gamma\over 2} {\cal V}\left( s \right) 
- \Omega {\cal W} \left( s \right) \\
 \dot{{\cal W}}\left(s\right) =   
\Omega {\cal V}\left( s \right)-{\Gamma\over 2} \left( 1 + s \right)
 {\cal W}\left( s \right) 
- {\Gamma \over 2} \left( 1 + s\right) {\cal Y} \left( s \right) \\ 
\dot{{\cal Y} } \left( s \right) = - { \Gamma \over 2} \left( 1 - s \right)
{\cal W} \left( s \right) - { \Gamma \over 2} \left( 1 - s \right) {\cal Y} \left( s \right).
\end{array}
\label{eqA01}
\end{equation}
These equations are 
exact within the rotating wave approximation and optical
Bloch equation formalism.  
In Eq. (\ref{eqA01}) $\Gamma$ is the spontaneous emission rate of
the electronic transition and
$\Omega$ is the Rabi frequency.
The time evolving detuning  is  
$\delta_L(t) = \omega_L - \omega_0 - \Delta\omega(t)$,
where $\omega_L$ $(\omega_0)$ is the laser frequency
(the molecule's bare frequency),
and $\Delta\omega(t)$ is the stochastic spectral diffusion process.
 The information
about the photon statistics is contained in
the moment generating function \cite{Brown}
\begin{equation}
2 { \cal Y } (s) \equiv
 \sum_{N=0}^\infty s^N P_N(t) 
\label{eq02}
\end{equation}
where  $P_N(t)$ is the probability of $N$ emission events
in the time interval $(0,t)$. Thus the solution of Eq. (\ref{eqA01})
yields the moments of $N$
\begin{equation}
\begin{array}{c c}
\overline{N(t) } = 2  {\cal Y}' (1) \ \ \ \ \
\overline{N^2(t)} = 2 {\cal Y}'' (1)
+ 2 {\cal Y}' (1),
\end{array}
\label{eqA02}
\end{equation}
with which the Mandel $Q$ parameter can in principle be obtained.
In Eq. (\ref{eqA02}),
and in what follows, 
we use the notation 
${\partial \over \partial s} g(s)|_{s=1} \equiv g'(1)$ 
and similarly for second order derivatives with respect to 
$s$.
The physical meaning of 
${\cal U}(s)$, ${\cal V}(s)$, 
and ${\cal W}(s)$ and their relation to
the standard Bloch equation was given in \cite{Brown}, and in
closely related work by Mukamel \cite{Mukamel}, some discussion on
this issue will follow Eq.    
(\ref{eqMT}).
Note that when $s \to 1$
the damping terms in Eq. (\ref{eqA01}) become small
[i.e. the $(1-s) \Gamma /2$ terms], hence relaxation of 
the generalized Bloch equations in the important limit of
$s \to 1$ is slow. 

 In what follows we will consider the
moments $\overline{N(t) }$, $\overline{N^2(t)}$. For this aim it
is useful to derive equations of motion for the
vector   
$z = \left\{ 
 {\cal U} (1), {\cal V}(1),{\cal W}(1),
{\cal Y}(1),{\cal U}'(1),
{\cal V}'(1),{\cal W}'(1),{\cal Y}'(1),{\cal Y}''(1) \right\}$.  
 Taking the first and the second derivative of  Eq. 
(\ref{eqA01}) with respect to $s$ and setting $s=1$, we find  
\begin{equation}
\dot{z} = M(t) z
\label{eqmtz}
\end{equation}
where $M(t)$ is a $9\times 9$ matrix
$M(t) =$ 
\begin{equation}
\left(
\begin{array}{c c c c c c c c c}
-{\Gamma\over 2} & \delta_L (t) &  0  &  0  &  0  &  0  &  0  &  0  &  0 \\
-\delta_L(t) & -{\Gamma \over 2} & - \Omega&0  &  0  &  0  &  0  &  0  &  0 \\
0& \Omega & - \Gamma  & - \Gamma &    0  &  0  &  0  &  0  &  0  \\     
0 & 0 & 0  & 0  &  0  &  0  &  0  &  0 & 0  \\                             
0 & 0 & 0  & 0  & - {\Gamma \over 2} & \delta_L (t) & 0  &  0 & 0 \\  
0 & 0 & 0  & 0  & - \delta_L (t)  & - {\Gamma \over 2} &-\Omega & 0 & 0 \\
0 & 0 & -{\Gamma \over 2} &- {\Gamma \over 2} & 0 & \Omega &-\Gamma & -\Gamma & 0 \\                             
0 & 0 & {\Gamma \over 2} & {\Gamma \over 2} & 0 & 0 & 0 & 0 & 0 \\    
0 & 0 &  0 & 0 & 0 & 0 & \Gamma & \Gamma & 0                              
\end{array}
\right)
\label{eqMT}
\end{equation}
The first three lines of $M(t)$ describe
the evolution of 
${\cal U} (1), {\cal V}(1),{\cal W}(1)$, these are the standard
optical Bloch equations in the rotating wave approximation.
The fourth line of $M(t)$ is zero, it yields 
$\dot{\cal{Y}}(1)=0$, this equation describes the normalization
condition of the problem namely 
${\cal Y}(1)=1/2$ for all times $t$ [to see this
use Eq. (\ref{eq02}) and $\sum_{N=0}^\infty P_N(t) = 1$].
The evolution of the remaining terms
${\cal U}'(1),
{\cal V}'(1),{\cal W}'(1),{\cal Y}'(1),{\cal Y}''(1)$ are of
current  interest since they describe the fluctuation of the
photon emission process. 
Solutions of time dependent equations
like Eq.
(\ref{eqmtz}) are generally extremely difficult to 
obtain, a formal solution is given in terms of
the time ordering operator $T$,
$z(t) = T \exp [\int_0^t M(t) {\rm d} t] z(0)$.
 
Eq. (\ref{eqmtz}) yields a general method for the
calculation of $Q$ for a single molecule undergoing
a spectral diffusion process. 
The aim of this Letter is to obtain an exact solution 
of the problem for an important stochastic process
used by Kubo and Anderson to investigate characteristic
behaviors of line shapes.
We assume $\Delta \omega(t) = \nu h(t)$ where $\nu$ describe
frequency shifts, and $h(t)$ describes a random telegraph process:
$h(t)=1$ or $h(t)=-1$. The transition rate between state
up (+) and state down (-) and vice versa is $R$.
This stochastic process was used to describe
behavior of line shapes in many systems \cite{Geva,Berne},
here our aim is
to calculate $Q$ describing the line shape fluctuations. 

\begin{figure}
\begin{center}
\epsfxsize=70mm
\epsfbox{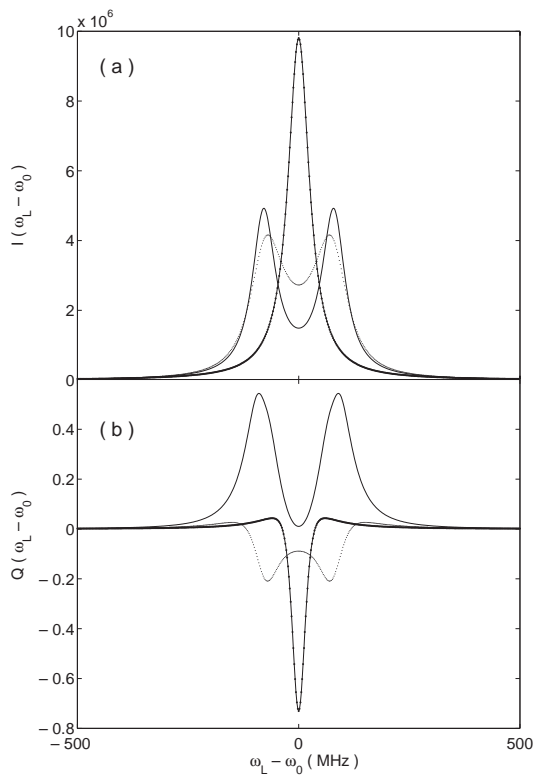}
\end{center}
\caption
{ In (a) we show the line shape $I(\omega_L)$ vs $\omega_L$ for
$\Gamma= 40$ MHz, $ \Omega = \Gamma / \sqrt{2}$, and $\nu = 2 \Gamma$.
   The line exhibits the well known 
transition between  fast ($R= 100 \Gamma$  dotted solid curve)
 and slow modulation limits ($R= \Gamma/10$ solid curve).
In (b) we show the $Q$ parameter, using the same parameters.
Three type of behaviors are observed: (i) for fast modulation
the $Q$ parameter attains its minimum on zero detuning. 
(ii) For an intermediate modulation,
$R = \Gamma/2$ (dotted curve) we obtain $Q<0$ however now the minimum of
$Q$ is on $\omega_L - \omega_0 = \pm \nu$. (iii) For the slow
process $Q$ exhibits super-Poissonian statistics $Q> 0$.
 Note that we
used $\Omega = \Gamma/ \sqrt{2}$ in order to maximize the absolute
value of $Q$ in the fast modulation limit.  
}
\label{fig1}
\end{figure}

 We use Burshtein's method \cite{Burshtein,Shore}
of marginal averages,
to solve the stochastic differential matrix equation
(\ref{eqmtz}). The method yields the average 
$\langle z \rangle$ with respect to the stochastic
process. We will calculate $\langle z \rangle$ in the limit
of long times, and then obtain the steady state behavior
of  the line shape 
$I\left( \omega_L \right)=\lim_{t \to \infty} {{\rm d} \over {\rm d} t}\langle \overline{N} (t) \rangle$ 
and the Mandel $Q$ parameter.
Let $\langle z \rangle_{\pm}$ be the average
of $z(t)$ under the condition that at time 
$t$ the value of $h(t) = \pm 1$ respectively. 
 $\langle z \rangle_{\pm}$ are called marginal averages,
the complete average is $\langle z \rangle = \langle z \rangle_{+}
+ \langle z \rangle_{-} $. 
The equation of motion for the marginal averages is 
\begin{equation}
\left(
\begin{array}{l}
\langle \dot{z}_{+} \rangle \\ 
\langle \dot{z}_{-} \rangle  
\end{array}
\right)=
\left(
\begin{array}{c c}
M_{+} - R I & R I  \\ 
RI           & M_{-} - R I  \\ 
\end{array}
\right)
\left(
\begin{array}{l}
\langle z_{+} \rangle \\ 
\langle z_{-} \rangle  
\end{array}
\right).
\label{eqMAT18}
\end{equation}
In Eq. (\ref{eqMAT18})
the matrix $M_{\pm}$ is identical to matrix $M$
in Eq. 
(\ref{eqMT})
when $\delta_L(t)$ is replaced by 
$\delta_L ^{\pm} =\omega_L - \omega_0 \mp \nu$,
 and $I$ is a $9\times 9$ identity matrix.
The $18 \times 18$ matrix in Eq. (\ref{eqMAT18}) is traceless, hence the 
time dependent solution of Eq. (\ref{eqMAT18}) 
is not obtained by direct inversion of it.
We found the long time behavior of Eq. (\ref{eqMAT18})
after a long journey which involved the following
five steps.
${\bf 1. }$  We obtained the steady state solution for
$\langle \{ {\cal U}(1)_{\pm}, {\cal V}(1)_{\pm}, {\cal W}(1)_{\pm},
{\cal Y}(1)_{\pm},{\cal Y}'(1)_{\pm}  \} \rangle$. 
${\bf 2. }$ We used the solutions
obtained in previous step to obtain inhomogeneous equations
for 
$x= \langle \{  
 {\cal U}'(1)_{+}, {\cal V}'(1)_{+},{\cal W}(1)_{+},
 {\cal U}'(1)_{-}, {\cal V}'(1)_{-},{\cal W}'(1)_{-}
\} \rangle$
\begin{equation}
\dot{x} = A x + b(t) 
\label{eqXAB}
\end{equation}
where 
\begin{equation}
A=\left(
\begin{array}{c c c c c c }
-{\Gamma\over 2} - R  & \delta^{+} _L &  0  &  R  &  0  &  0   \\
-\delta^{+}  _L & -{\Gamma \over 2} -R  & - \Omega &0  &  R  &  0  \\
0& \Omega & - \Gamma - R   & 0  &    0  &  R   \\     
R & 0 & 0  &  -{\Gamma \over 2} -R  &  \delta^{-} _L   &  0  \\
0 & R & 0  & - \delta^{-} _L & - {\Gamma \over 2} - R & -\Omega   \\  
0 & 0 & R  & 0  & \Omega & - \Gamma - R 
\end{array}
\right)
\end{equation}
and $b(t) = ( 0, 0, b_{+}(t) , 0, 0, b_{-}(t) )$ with
\begin{widetext}
\begin{equation}
 b_{\pm}(t) = -{\Gamma \over 8} - {\Gamma \over 2} {\cal W}^{ss}(1)_{\pm}-
{\Gamma^2 t \over 8} \left\{
1 + 2 \left[ {\cal W}^{ss}(1)_{+} + {\cal W}^{ss}(1)_{-} \right] \right\}
 \pm { \Gamma^2 \over 8 R } \left[ {\cal W}^{ss}(1)_{-} - {\cal W}^{ss}(1)_{+} \right]. 
\label{eqBB}
\end{equation}
and
\begin{equation}
{\cal W}^{ss}(1)_{+}  = {\Gamma \over 4} \left[\left(A^{-1}\right)_{33} + \left(A^{-1}\right)_{36}\right] \ \  \ \ \ \ \ \ \ {\cal W}^{ss}(1)_{-} ={\Gamma \over 4}  \left[\left(A^{-1}\right)_{63} + \left(A^{-1}\right)_{66}\right].
\end{equation}
We note that ${\cal W}^{ss}(1)_{\pm}$ yields the steady state (ss) marginal
averages of the population difference between the upper
excited state and ground state.
${\bf 3. }$  Using Mathematica we found an analytical expression
for $A^{-1}$, the inverse of the Matrix $A$, in terms of the
parameters of the problem $R,\nu , \Gamma , \omega_L- \omega_0, \Omega$.
${\bf 4.}$ Using $A^{-1}$ and 
Eq. (\ref{eqXAB}) we find the long time behavior of
the  vector $x$. We then  use
Eq.
(\ref{eqMAT18})
to obtain steady state behavior of  
$\langle {\cal Y}''(1)^{{\rm ss}} \rangle_{\pm}$. 
${\bf 5.}$ Finally we obtain the $Q$ parameter using:
$Q=  \left[ \langle  {\cal Y}''(1)^{{\rm ss}} \rangle -2 \langle 
 {\cal Y}'(1)^{{\rm ss}} \rangle^2 \right] / \langle {\cal Y}'(1)^{{\rm ss}} \rangle$.

Following these steps we obtain the main result of this Letter
%
%\begin{widetext}
\begin{equation}
 Q =  
 {\Gamma^2 \over 2} \sum_{i=3,6} \sum_{j=3,6} \left( A^{-1}A^{-1} \right)_{ij}
+ {\Gamma^2 \over R I (\omega_L) }\left[ {\cal W}^{ss}(1)_{+} - {\cal W}^{ss}(1)_{-} \right]^2 + 
{\Gamma \over I\left( \omega_L \right)}
\sum_{k=\pm}  \left[ {\cal W}^{ss}(1)_{k} \left( 1 +
4 {\cal W}^{ss}(1)_{k} \right) \right], 
\label{eqMain}
\end{equation}
\end{widetext}
where the line 
is
\begin{equation}
I\left(\omega_L \right) = \Gamma\left[ { 1 \over 2} + {\cal W}^{ss}(1)_{+}  
+{\cal W}^{ss}(1)_{-} \right]. 
\label{eqLine}
\end{equation}
The general expression for
$Q$ in terms of original parameters of the problem
can be found in \cite{remark}. Below we will
analyze the physical behaviors of $Q$
obtaining simple equations in limiting cases.

 In Fig. \ref{fig1} we show the line shape and the $Q$ parameter.
We choose $\Omega=\Gamma/\sqrt{2}$, fix $\nu$, and vary the rate $R$.
The line exhibits two well known behaviors: a fast
modulation limit where the line is Lorentzian and a slow modulation
limit where the line exhibits two peaks on $ \omega_L = \pm \nu$. The $Q$
parameter exhibits three behaviors: (i) fast modulation limit,
where the single molecule exhibits a sub-Poissonian
behavior, and $Q$ attains its minimum on zero detuning.
(ii) Intermediate modulation limit where photon statistics
is still sub-Poissonian, however now $Q$ attains its minimum
on $ \omega_L = \pm \nu$. And (iii) a slow modulation limit where
a super-Poissonian behavior is observed. 
We now analyze these limits in some detail.
Without loss of generality we use units where $\Gamma=1$,
and set $\omega_0=0$ hence $\omega_L$ is the detuning frequency.

For zero detuning, $\omega_L = 0$,  we obtain
from Eq. (\ref{eqMain})
\begin{equation}
Q = 
-{2\,{\Omega}^2\,\left[ 4\,{\nu}^2\,\left( -1 + 4\,R \right)  +
      3\,\left( 1 + 4\,R \right)^2 \right] \over \left[ 4\,{\nu}^2 +
       \left( 1 + 2\Omega^2 \right) \,\left( 1 + 4\,R \right)  
\right]^2}.
\label{refzdet}
\end{equation}
Eq. (\ref{refzdet}) indicates a transition
from sub-Poissonian statistics $(Q<0)$ to super-Poissonian statistics
$(Q>0)$ when  
$4\nu^2=3(1+4 R)^2/(1-4 R)$. For fast processes satisfying $R>1/4$,
(i.e. $R > \Gamma/4$ in natural units)
we find sub-Poissonian behavior for all
values of $\nu$ and $\Omega$. 
As mentioned in introduction an important issue is how to choose
$\Omega$ in such a way to minimize $Q$ in the sub-Poissonian
regime (i.e., how to maximize the quantum nature of
the photon statistics).  Using Eq. (\ref{refzdet}) 
it is easy to find the lower bound $Q \ge - 3/4$.
The absolute  minimum $Q=-3/4$ is found when $\Omega= 1/\sqrt{2}$ and
$\nu =0$, namely for a 
stable molecule whose absorption frequency is not fluctuating.
For fixed $\nu$ and $R$ the largest quantum fluctuations are
observed
when
$\Omega_{{\rm min}}^2 = ( 1 + 4 \nu^2 + 4 R)/ [2 ( 1 + 4 R)]$. 
Note however that for fast processes
$\nu<<R$, the parameters $\nu$ and $R$
cannot be directly deduced in
experiment, due to motional narrowing effect. In Eq. 
(\ref{eqEFF})
we will give an 
expression for $\Omega_{{\rm min}}$ in terms of physical
observables. 

 Consider the fast modulation limit.
If we take $R \to \infty$ keeping $\Omega$ and $\nu$ fixed we obtain
from Eq. (\ref{eqMain})
\begin{equation}
\lim_{R \to \infty} Q= -  { 2 \Omega^2 \left(  3 - 4 \omega_L ^2 \right) \over
\left( 1 + 2 \Omega^2 + 4 \omega_L ^2 \right)^2 }.
\label{eqMand}
\end{equation}
This result was derived by  Mandel for an atomic transition in the
{\em absence} of spectral transition. As expected 
when the time scale for spectral jumps becomes very short,
the molecule does not respond to the stochastic fluctuation.  
A more physically interesting case is to let $R \to \infty$ 
and $\nu \to \infty$
but keep the spectral diffusion (SD) contribution to the 
line width $\Gamma_{{\rm SD}} \equiv \nu^2 / R$ remain finite.
In this limit, the line shape is Lorentzian  
and it exhibits motional narrowing
(i.e. as $R$ is increased the line becomes narrower). 
 In this
fast limit we obtain from Eq. (\ref{eqMain})
\begin{equation}
Q_{{\rm fast}} = 
- { 2 \Omega^2 \left[ 3 + 5 \Gamma_{{\rm SD}}^2 + \Gamma_{{\rm SD}}^3 - 4 \omega_L\ ^2 + \Gamma_{{\rm SD}} \left( 7 + 4 \omega_L ^2 \right) \right]
\over
\left[ 1 + \Gamma_{{\rm SD}}\ ^2 + 2 \Omega^2 + 2 \Gamma_{{\rm SD}} \left( 1 + \Omega^2 \right) + 4 \omega_L ^2 \right]^2 }.
\end{equation}
Note that photon statistics are sub-Poissonian provided that
the detuning is not too large. The $Q$ parameter obtains its
minimum when
\begin{equation}
\Omega^2 _{{\rm min}} = { \left ( 1 + \Gamma_{{\rm SD}} \right)^2 + 4 \omega_L ^2 
\over
2 \left( 1 + \Gamma_{{\rm SD}} \right) },
\label{eqEFF}
\end{equation}
and for zero detuning
$Q_{{\rm min}} =  - (\Gamma_{{\rm SD}} + 3 ) / 4 ( 1 + \Gamma_{ { \rm SD}} )$.
The simple Eq. (\ref{eqEFF})
is important, since it shows how to choose
the Rabi frequency in order to obtain strong sub-Poissonian
behavior. Since $\Gamma_{{ \rm SD}}$ is a physical observable
(unlike $R$ and $\nu$) Eq. (\ref{eqEFF}) is of practical value.

  To investigate the intermediate and slow modulation limit
we consider the value of $Q$ for $ \omega_L = \nu$.
We consider a strong coupling limit $ \nu >> 1$,
(e.g. spectral jumps of the order
of $\nu=\mbox{GHz} >> \Gamma$). 
From Eq. (\ref{eqMain}) we obtain
$$ \lim_{\nu \to \infty} Q_{\omega_L=\nu} = $$
\begin{equation}
-{ \left( 1 + 2 R \right) \Omega^2 \left[ - 1 + 16 R^2 + 8 R^3 - 2 \Omega^2 + 
2 R \left( 2 + \Omega^2 \right) \right] 
\over 2 R \left[ 1 + 4 R^2 + 2 \Omega^2 + 2 R \left( 2 + \Omega^2 \right) \right]^2 }.
\label{eqwwww}
\end{equation}
It is easy to see that 
Eq.
(\ref{eqwwww}) exhibits both sub-Poissonian
and super--Poissonian behaviors.
When the process is very {\em slow}, namely $R \to 0$, we obtain
\begin{equation}
\lim_{\nu \to \infty}
Q_{ \omega_L = \nu }\sim {\Omega^2 \over  2 R \left( 1 + 2 \Omega^2 \right) } 
\end{equation}
a super-Poissonian behavior. 
In the {\em intermediate modulation limit},
for example when $R = 1$, we obtain  
\begin{equation}
 \lim_{\nu \to \infty}
 Q_{\omega_L = \nu} = - { 81 \Omega^2 \over  2  \left( 9 + 4 \Omega^2 \right)^2 }
\end{equation}
a sub-Poissonian behavior.
In this case the fluctuations are strongest when $\Omega^2 _{\min} = 9/4$
and then $Q_{{\rm min}} = - 9/32$. It is important to notice that
in the same limit $ \nu \to \infty$ we have for zero
detuning $|Q| \propto \nu^{-2} \rightarrow 0$ 
hence in the slow and intermediate
modulation limits the fluctuations on
$\omega_L = \pm \nu$ are much stronger than the fluctuations
on zero detuning (an example is in Fig.
\ref{fig1}).  

The detailed interaction between a chromophore
and a bath is usually extremely complex, still for many fast processes
line shapes are simple Lorentzians (universality). 
We expect that certain behaviors
of $Q$ obtained in this work are also universal, mainly the behavior in the
fast modulation limit and the transition between fast, intermediate, and slow
limits.
We also expect, but have no proof,
that the maximum of the quantum fluctuations, is attained 
close to  $\Omega_{{\rm min}} ^2 / \Gamma = (\Gamma_{{\rm SD}} + \Gamma)/2$
[i.e., Eq. (\ref{eqEFF}) in natural units and for zero detuning],
also in other models of line broadening. 
Finally, the optimization
of external fields 
to obtain strong sub-Poissonian and anti-bunching
signals from single molecules is a new topic of research
worthy of further investigation.

\newpage

$$ $$
$$ $$
\begin{widetext}
\section{Exact Solution}

The exact solution for the $Q$ parameter, with $\Gamma=1$, in the long
time limit, $ Q = \mbox{Numerator}[Q]/\mbox{Denominator}[Q] $
is
$$\mbox{Denominator}[Q] = $$
\begin{equation}
\begin{array}{l}
R\,\left( 4 {\omega_L}^2\,\left( 1 + 2\,R \right)  + 
     4\,{\nu}^2\,\left( 1 + 6\,R + 8\,R^2 \right) \right.  + 
 \left.  \left( 1 + 4\,R \right) \,\left( 1 + 6\,R + 8\,R^2 + 2\,\Omega^2 \right) 
     \right) 
 \,\left( 16\,{\nu}^4\,\left( 1 + 2\,R \right) \right.  + \\
       8\,{\nu}^2\,\left( 1 - 4\,{\omega_L}^2\,\left( 1 + 2\,R \right)  + 
          2\,\Omega^2 + 2\,R\, \left( 3 + 4 R \right)\left( 1 + \Omega^2 \right)   
         \right)  + 
       \left( 1 + 4\,{\omega_L}^2 + 2\,\Omega^2 \right)  \,\\
 \left.       \left(4  {\omega_L}^2\,\left( 1 + 2\,R \right)  + 
          \left( 1 + 4\,R \right) \,\left( 1 + 6\,R + 8\,R^2 + 2\,\Omega^2 \right) 
          \right)  \right)^2
\end{array}
\end{equation}

$$\mbox{Numerator}[Q] = $$
\begin{equation}
\begin{array}{l}
 -2\,\Omega^2\,\left( 256\,{\nu}^8\,R\,{\left( 1 + 2\,R \right) }^3\,
      \left( -1 + 16\,R^2 \right)  -  
     \left( -3 + 4\,\omega_L^2 \right) \,R\,
      \left( {\omega_L}^2\,\left( 4 + 8\,R \right)  + 
          \left( 1 + 4\,R \right) \,\left( 1 + 6\,R + 8\,R^2 + 2\,\Omega^2 \right) 
          \right)^3 \right.  \\
 - 128\,{\nu}^6\,\left( 1 + 2\,R \right)^2\, 
 \left( 4\,{\omega_L}^2\,\left( 1 + 4\,R + 4\,R^2 + 16\,R^3 + 32\,R^4
           \right)  - 3\,R\,\left( 1 + 4\,R \right) \, 
       \left( 24\,R^2 + 32\,R^3 - \Omega^2 + 4\,R\,\left( 1 + \Omega^2 \right) 
           \right)  \right) \\
 + 32\,{\nu}^4\,\left( 1 + 2\,R \right) \,
      \left( 16\,{\omega_L}^4\,{\left( 1 + 2\,R \right) }^2\, 
         \left( 2 + 5\,R + 8\,R^3 \right)  + 
        4\,{\omega_L}^2\,\left( 1 + 6\,R + 8\,R^2 \right) \, 
         \left( -2 + 112\,R^4 + 64\,R^5 - 4\,\Omega^2 +  \right. \right. \\
            \left.  R^2\,\left( 76 - 8\,\Omega^2 \right)
- R\,\left( -4 + \Omega^2 \right)  + 
           16\,R^3\,\left( 10 + \Omega^2 \right)  \right)  + 
        3\,R\,\left( 1 + 4\,R \right) \, 
         \left( 1 + 832\,R^4 + 512\,R^5 + \Omega^2 - 2\,\Omega^4 + 
           32\,R^3\,\left( 16 + 5\,\Omega^2 \right) + \right. \\
          \left.  \left. 4\,R^2\,\left( 37 + 32\,\Omega^2 \right)  + 
           R\,\left( 20 + 26\,\Omega^2 + 8\,\Omega^4 \right)  \right)  \right)  -  
     8\,{\nu}^2\,\left( 64\,{\omega_L}^6\,\left( 1 + 2\,R \right)^4 - 
        R\,\left( 1 + 4\,R \right) \,  
         {\left( 1 + 6\,R + 8\,R^2 + 2\,\Omega^2 \right) }^2\,
         \left( 4 + 152\,R^2 \right. \right.  \\
 + 160\,R^3 - \Omega^2 + 
          \left. 4\,R\,\left( 11 + \Omega^2 \right)  \right)  - 
        16\,{\omega_L}^4\,{\left( 1 + 2\,R \right) }^2\, 
         \left( -2 + 40\,R^3 - 4\,\Omega^2 - 4\,R^2\,\left( -5 + 6\,\Omega^2 \right)  - 
           R\,\left( 4 + 17\,\Omega^2 \right)  \right)  - 
        4\,{\omega_L}^2\, \\
\left( 1 + 2\,R \right) \, 
         \left( 7680\,R^6 + 2048\,R^7 + 
256\,R^5\,\left( 37 + \Omega^2 \right)  - 
           {\left( 1 + 2\,\Omega^2 \right) }^2 + 
           32\,R^4\,\left( 169 + 36\,\Omega^2 \right)  +   
          8 R^3\,\left( 191 + 108\,\Omega^2 - 8\,\Omega^4 \right)  \right.+ \\
 \left. \left.  \left.  4 R^2\,\left( 47 + 43 \,\Omega^2 - 12 \,\Omega^4 \right)  - 
           2\,R\,\left( -1 + 5\,\Omega^2 + 14\,\Omega^4 \right)  \right)  \right) 
     \right) .

\end{array}
\end{equation}
The line shape  $(\Gamma=1)$ is
$$ I\left( \omega_L  \right) = $$
\begin{equation}
\begin{array}{c}
\left[ \Omega^2\,\left( 1 + 32\,R^2 + 32\,R^3 + 
4\,\nu^2\,\left( 1 + 6\,R + 8\,R^2 \right)  + 2\,\Omega^2 + 4\,\omega_L ^2 + 
2\,R\,\left( 5 + 4\,\Omega^2 + 4\,\omega_L ^2 \right)  \right)  \right] / 
\\
\left[ 16\,\nu^4\,
      \left( 1 + 2\,R \right)  + 
     8\,\nu^2\,\left( 1 + 2\,\Omega^2 + 8\,R^2\,\left( 1 + \Omega^2 \right)  - 4\,\omega_L ^2 + 
        R\,\left( 6 + 6\,\Omega^2 - 8\,\omega_L ^2 \right)  \right) \right. + \\
 \left.  \left( 1 + 2\,\Omega^2 + 4\,\omega_L ^2 \right) \,
      \left( 1 + 32\,R^2 + 32\,R^3 + 2\,\Omega^2 + 4\,\omega_L ^2 + 
        2\,R\,\left( 5 + 4\,\Omega^2 + 4\,\omega_L ^2 \right)  \right) \right]
\end{array}
\end{equation}

\end{widetext}


\begin{thebibliography}{99}

\bibitem{MO}  W. E. Moerner and M. Orrit,
{\em Science} {\bf 283} 1670 (1999). 

% Nanosecond dynamics of single polypeptide molecules
% revealed by photoemission statistics
% of fluoresence eneregy transfer: A theoretical study.
%\bibitem{WangM} Z. S. Wang, D. E. Makarov  {\em J. Phys. Chem. B }
% {\bf 107} 5617 (2003)

\bibitem{MM} D. E. Makarov, H. Metiu {\em J. Chem. Phys.}
{\bf  115} 5989--5993 (2001)
 
\bibitem{Schenter} G. K. Schenter, H. P. Lu, X. S.  Xie
{\em J. Phys. Chem. A}  {\bf 103} 10477 (1999).

\bibitem{Berez} A. M. Berezhkovskii, A. Szabo, and G. H. Weiss
{\em J. Phys. Chem. B} {\bf 104} 3776 (2000).

\bibitem{Barsegov2} V. Barsegov, S. Mukamel 
{\em J. of Chem. Phys.}  {\bf 116} 9802 (2002).

% in low temperature glasses
\bibitem{Geva}  E. Geva, and J. L. Skinner,
{\em J.  Phys.  Chem. B}  {\bf 101},  8920 (1997).

%{\em Time-Dependent Fluctuations in Single Molecule Spectroscopy: A Generalized%Wiener--Khintchine Approach}
\bibitem{PRL} E. Barkai, Y. Jung, R. Silbey  
{\em Phys. Rev. Lett.}  {\bf 87} 207403 (2001)

% A Stochastic Theory of Single Molecule Spectroscopy
\bibitem{ADV} Y. Jung, E. Barkai, and R. Silbey {\em Adv. Chem. Physics} {\bf 123}, 199 (2002) and cond-mat/0311428


\bibitem{Wolf} L. Mandel, and E. Wolf,
{\em Optical Coherence and Quantum Optics}
(Cambridge University Press, New York, 1995).  

\bibitem{Plenio} M. B. Plenio, and P. L.  Knight
{\em Rev. of  Mod. Phys.}  70:101--144 (1998).

\bibitem{Mandel1} L. Mandel, {\em Opt. Lett.} {\bf 4} 205 (1979).


\bibitem{Short} R. Short, and  L. Mandel, {\em Phys. Rev. Lett.} {\bf 51} 384 (1983).

%{\em Photon anti--bunching in the Fluorescence of a Single Dye Molecule
% Trapped in a Solid}
\bibitem{Or} Th. Basche, W. E. Moerner, M. Orrit, and H. Talon,
{\em Phys. Rev.
Lett.} {\bf 69}, 1516  (1992).

%{\em Non-Classical Photon statistics in Single Molecule Fluorescence at
%Room Temperature}
\bibitem{Zu1} L. Fleury, J-M. Segura, G. Zumofen, B. Hecht, and U. P. Wild,
{\em Phys. Rev. Lett.} {\bf 84}, 1148 (2000).


% Single Molecule Photon Counting Statistics via Generalized
% Optical Bloch Equations
\bibitem{Brown}  Y. Zheng, F. L. H. Brown {\it Phys. Rev. Lett.}  {\bf 90}
238305  (2003)

% Photon Statistics; Nonlinear Spectroscopy of Single Quantom Systems
% Phys. Rev. A submitted
\bibitem{Mukamel} S. Mukamel, cond-mat/0307768


\bibitem{Burshtein} A. I. Burshtein {\em Sov. Phys. JETP} {\bf 22} 937 (1966).

\bibitem{Shore}  B. W. Shore {\em The Theory of Coherent Atomic Excitation} 
Vol. 2,  Wiley New-York (1990). 

\bibitem{remark} We suggest that exact expression for $Q$ will be published
as an EPAPS document, see http://www.aip.org./pubservs/epaps.html

\bibitem{Berne} B. J. Berne, and R. Pecora {\em Dynamic Light Scattering}
Dover Publication, Chapter 6.

% Spectral  Diffusion of Single Molecule Fluorescence: A
% Probe of Low-Frequency Localized Excitations in Disordered
% Solids.
%\bibitem{Reilly} P. H. Reilly, and J. L. Skinner, {\em Phys. Rev. Lett.}
%{\bf 71} 4257 (1993)

\end{thebibliography}
\end{document}